\documentclass[letterpaper,twocolumn,10pt]{article}
\usepackage{usenix-2020-09}

\usepackage{tikz}
\usetikzlibrary{positioning, arrows, decorations.pathreplacing, calc}
\usepackage{amsmath}

\usepackage{xspace}
\usepackage{graphicx}
\usepackage{subcaption}
\usepackage{array}        
\usepackage{multirow}     
\usepackage{booktabs}     
\usepackage{caption}      
\usepackage{geometry}     
\usepackage{graphicx}
\usepackage{subcaption}
\usepackage{array}        
\usepackage{multirow}     
\usepackage{booktabs}     
\usepackage{caption}      
\usepackage{geometry}     
\usepackage[acronym]{glossaries}
\glsdisablehyper

\newacronym{tcb}{TCB}{Trusted Computing Base}
\newacronym{mmu}{MMU}{Memory Management Unit}
\newacronym{mpk}{MPK}{Memory Protection Keys} 
\newacronym{poe}{POE}{Permission Overlay Extension}
\newacronym{el0}{EL0}{Exception Level 0}
\newacronym{jit}{JIT}{Just in-Time}
\newacronym{irt}{IRT}{Instruction Region Table}
\newacronym{dpot}{DPOT}{Data Permission Overlay Table}
\newacronym{ttt}{TTT}{TIndex Transition Table}

\begin{document}

\date{}

\title{\Large \bf Complets: Universal Compartmentalisation and Programming Model For Arm Permission Overlay Extension 2}

\author{
{\rm Vasily A. Sartakov}\\
Huawei R\&D
} 

\maketitle

\begin{abstract}
Arm Permission Overlay Extension (POE) is an intra‑process isolation mechanism based on memory protection keys.
This mechanism partitions virtual memory into regions whose access permissions can be reconfigured without invoking higher exception levels.
Because POE does not enforce privilege separation, POE‑based protection domains are applicable only as a software fault‑isolation technique.

The POE2 is an isolation mechanism that substantially extends POE1.
Dealing with memory protection keys, it also introduces new registers and tables that regulate the operations permitted for code executing within protection domains associated with different keys.
A permission is determined by the \emph{spatial} index of the executing code (i.e., the protection key associated with code), the spatial index of the accessed memory, the operation type, and a new attribute of the thread context: its \emph{temporal} index.

POE2 provides stronger security guarantees than POE1, but also introduces considerable architectural complexity.
Effective permissions arise from the interaction of multiple hardware components, making direct programming non-trivial and prone to subtle errors.
In the paper, we present a detailed analysis of POE2 and introduce a universal programming model with a strong security model for POE2‑based systems.
The model abstracts the complexity of spatial and temporal indices while enabling typical patterns of partitioned software constructed using intra‑process isolation.

\end{abstract}

\section{Introduction}
Virtual memory and processes are fundamental mechanisms that enable isolated program execution.
Each program executes within its own address space, without visibility into how virtual addresses are backed by hardware or whether other programs exist.
If virtual address spaces are isolated and no pages are shared, programs cannot directly influence one another and may communicate only through the privileged intermediary -- the kernel.

The kernel serves as this privileged intermediary and provides mechanisms to configure the \gls{mmu}.
Interaction between processes requires transitions into privileged mode, which impose performance overhead on communicating programs.
Consequently, the trade-off between the size of the \gls{tcb} and communication performance is driven by practical considerations: secure deployments favour finer‑grained isolation with higher communication costs, whereas less secure deployments permit more monolithic software with lower overhead.

One approach to reducing the overhead associated with process‑based isolation -- commonly referred to as the \gls{mmu} tax -- is to employ intra‑process isolation mechanisms and construct communication primitives that avoid privileged transitions.
Over the last decade, several techniques have been proposed: Intel \gls{mpk}~\cite{intel_sdm_2018} and Arm \gls{poe}~\cite{arm_permission_indirection_overlay} allow assigning protection keys to memory pages and modifying page attributes without kernel involvement.
Once keys are assigned, the effective permissions of associated pages are determined by the key’s permissions, and updates to key attributes do not require kernel mediation.
Using these extensions, prior work has demonstrated fine‑grained software partitioning with low performance overhead~\cite{11023477}. 

A key limitation of Arm \gls{poe} and Intel \gls{mpk} is the absence of privilege separation within compartments -- software components isolated using protection keys.
With \gls{mpk}, compartmentalised code can modify access rights of other compartments via the \texttt{wrpkru} instruction~\cite{epk}.
On Arm, compartmentalised code can similarly modify the \texttt{POR\_ELx} register.
In both cases, these operations occur without kernel oversight, enabling efficient cross‑compartment communication but preventing the use of these mechanisms to isolate adversarial code.
Despite attempts to reduce the likelihood of executing security‑sensitive instructions~\cite{gu2020harmonizing, hodor, erim}, protection‑key‑based compartment isolation has been treated primarily as software‑fault isolation.

\begin{figure}[t!]
    \centering
    \includegraphics[width=0.40\textwidth]{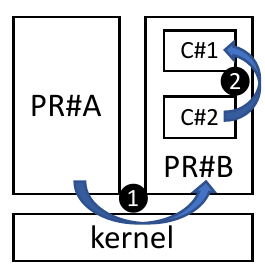}
    \caption{Communication between isolated parties: \textcircled{1} IPC between processes using kernel; \textcircled{2} Cross-compartment calls without the kernel }
    \label{fig:proc-comp}
\end{figure}

\newpage

Arm addressed the security limitations of \gls{poe}\footnote{Next and following, POE will stand either for the general term Permission Overlay Extension or refer specifically to POE1} and announced~\cite{arm_poe2_vmte_blog} a new version, \gls{poe}2.
\gls{poe}2 introduces privilege separation for compartmentalised code.
As in \gls{poe}1, the address space can be partitioned into protection domains by assigning keys, but permissions are now determined by a combination of indices associated with the executing code’s location, the thread state, the operation type, and the target memory.

Programming \gls{poe}2 requires kernel involvement, but -- as with \gls{poe} and \gls{mpk} -- transitions between isolated components can occur without kernel mediation.
However, the permissions governing operations are defined in tables managed by the operating system.
Arm does not provide an explicit programming model for \gls{poe}2, and in this work we present a programming model that abstracts the complexity of new instructions, tables, and states while offering universal primitives for implementing typical intra‑process isolation patterns.

The paper makes the following contribution: It
\begin{enumerate}
	\item provides an overview of \gls{poe}2 for \gls{el0}, including key mechanisms, new instructions, registers, and tables
	\item analyses the security properties of \gls{poe}2 and shows that secure transitions between isolated components require trusted elements
	\item introduces a universal programming model that implements key intra‑process design patterns, including asymmetrical (\gls{jit} isolation), symmetrical (mutually distrusted parties), and enclave‑style reversed-sandbox.
\end{enumerate}

The rest of the paper is organised as follows:
We begin with background on \gls{poe}2 and its security analysis in Section~\ref{sec:bg}.
In Section~\ref{sec:program_model}, we introduce the \gls{poe}2 programming model, which abstracts its complexity and enables the implementation of typical design patterns, considered later in Section~\ref{sec:impl}.

\section{Background}\label{sec:bg}
\gls{poe}2 introduces a set of new instructions, registers, and in‑memory tables.
It builds on prior extensions, including \gls{poe}1 and the Permission Indirection Extension (PIE).
Different exception levels may employ different mechanisms for transitions between software components: for exception level 0 (user space) a single approach is available, and this work focuses on that case.

\gls{poe}1 comprised two principal elements: POIndexes (keys) stored in page table entries, and the \texttt{POR\_ELx} registers, whose bit fields specify permissions applied to memory associated with a given key.
\gls{poe}2 reuses and extends these mechanisms: POIndexes remain encoded in page table entries, and \gls{poe}1 may be used together with, or independently of, \gls{poe}2.
\gls{poe} permissions are no held in a single register, instead they reside in memory tables such as \gls{irt}, \gls{ttt}, \gls{dpot}, and others.

Permissions are no longer determined solely by the \emph{spatial} characteristic of the CPU context, i.e. the key associated with the currently executing code or the accessed memory.
A new register, \texttt{TINDEX}, introduces a \emph{temporal} characteristic of a CPU context, providing an orthogonal dimension for identifying contexts: a thread may switch from one \emph{state} to another by changing this index, and the permitted operations are defined by the combination of the spatial and temporal characteristics of the context, the operation type (fetch, read, write), and the spatial characteristics of the target memory~\cite{arm_aarch64_s1poe2_2025}.
In other words, for non‑\gls{poe}2 instructions (for example load, store, fetch), \textbf{the combination of spatial and temporal indexes defines the current unique index of the executing thread, and the effective permissions applied to the current instruction are determined by the combination of this index and the spatial index of the target memory.}

\begin{figure*}[t]
    \centering

    \begin{subfigure}{0.49\textwidth}
	\centering
	\includegraphics[width=0.7\textwidth]{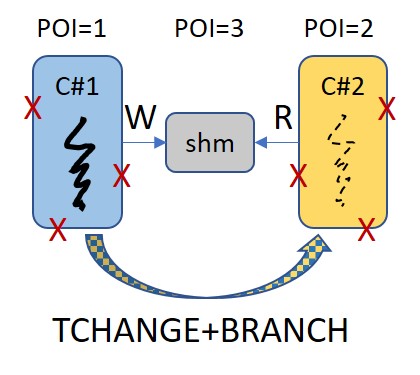}
	\caption{Two compartments with shared buffer and transition}
	\label{fig:tchange-br}
\end{subfigure}
    \begin{subfigure}{0.49\textwidth}
        \centering
        \includegraphics[width=\textwidth]{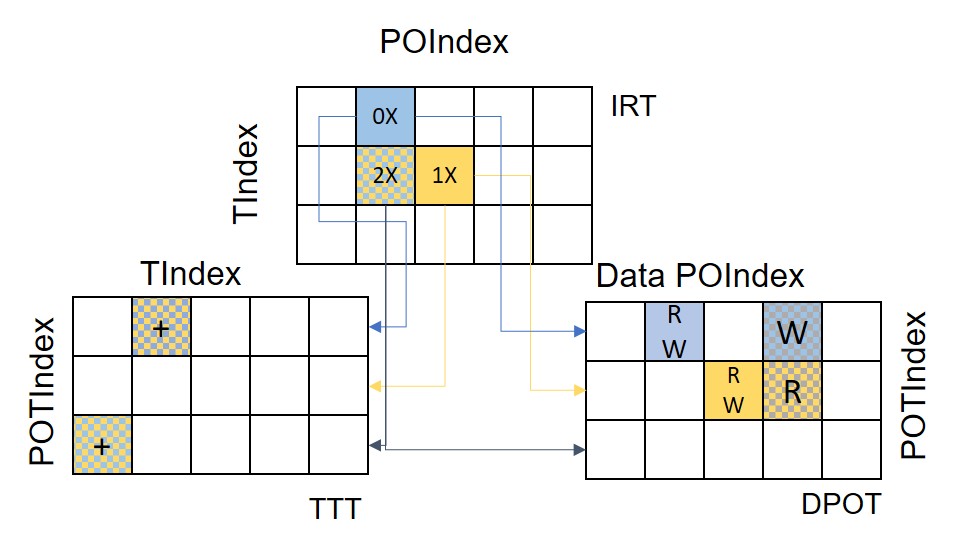}
        \caption{Key tables defining permissions: IRT, DPOT, TTT}
        \label{fig:tables}
    \end{subfigure}

    \caption{Two isolated yet components and the corresponding \gls{poe}2 configuration: Component \texttt{C\#1} with POIndex=1 can write into shared memory \texttt{shm} with POIndex=3  and switch to Component \texttt{C\#2} (POIndex=2) with read-only access to shared memory, performing \texttt{TCHANGE} and \texttt{BR} instructions}
    \label{fig:2-general}
\end{figure*}

Figure~\ref{fig:2-general} illustrates a simple scenario with two interacting software components, and the corresponding \gls{poe}2 table configuration.
Isolated component \texttt{C\#1} has write access to the shared memory \texttt{shm}, while isolated component \texttt{C\#2} has read access.
Both \texttt{C\#1} and \texttt{C\#2} are mutually distrusted and lack access to data outside their own compartments and the shared buffer; however, \texttt{C\#1} is permitted to perform a domain transition to \texttt{C\#2}.

The corresponding \gls{poe}2 table configuration is shown in Figure~\ref{fig:tables}.
The \gls{irt} is indexed by POIndex (the key in the page tables entries associated with a page) and by TIndex -- the new register in the CPU context.
Each \gls{irt} entry combines two state indexes, POTIndex and FGDTIndex (not shown), with an execution flag that enables fetch in that state~\cite{arm_aarch64_s1poe2_2025}.
POTIndex indexes the \gls{dpot} and the \gls{ttt}, while FGDTIndex is used for fine‑grained dynamic traps.

To enable the scenario in Figure~\ref{fig:tchange-br}, the \gls{irt} must include three POTIndexes -- one for each party involved.
POTIndex=0 is used for the state TIndex=0 and POIndex=1.
This POTIndex corresponds to ordinary execution inside the compartment created with key 1.
Fetch is enabled for this index combination.

POTIndex=1 is used for the state TIndex=1 and POIndex=2 and defines execution of the second component.
The shared memory is not executable and therefore lacks the X bit, and it has no \gls{irt} entry for execution.
A transition state that enables switching between isolated binaries is represented by POTIndex=2, which is defined for TIndex=1 and POIndex=1.
This index represents an execution state of the isolated code with permissions distinct from ordinary execution states.

Isolated contexts in \texttt{C\#1} and \texttt{C\#2} possess distinct TIndex and POIndex values, which prevents direct code transitions between them.
Indeed, in a single step, only one index may change: a temporal index effected by the \texttt{TCHANGE} instruction, or a spatial index effected by a branch instruction (which transfers control to memory associated with a different POIndex).
The temporal and spatial indices are represented and modified independently, and there is no mechanism to update both indices atomically. Therefore, a single‑step transition that alters both indices is impossible.
POE2 nevertheless permits controlled transitions by composing these single‑index operations: a transition proceeds via an intermediate state in which one index retains its previous value, so the overall change is realised as a sequence of \texttt{TCHANGE} and branch operations.


POTIndex=2 describes the intermediate state in which code from \texttt{C\#1} is spatially associated with \texttt{C\#1} but has a TIndex equal to \texttt{C\#2}, enabling a branch into \texttt{C\#2}’s memory.
Changes to TIndex are governed by the \gls{ttt}, which is indexed by POTIndex and the target TIndex.
Consequently, isolated code may elevate its temporal index only to values for which transitions are explicitly enabled.
Branches are not similarly constrained as they are generally permitted between POIndexes and the same TIndex but can be restricted by using different TIndex values.
As a result, \texttt{C\#1} and \texttt{C\#2} have unique TIndex–POIndex pairs.

The \gls{ttt}, which controls transitions betweens TIndexes, contains two records: one for POTIndex=0 (with TIndex=0) permitting transition into TIndex=1 so that a branch into \texttt{C\#2} becomes possible, and one for POTIndex=2 (with TIndex=1) permitting transition back into TIndex=0.
Forward and return transitions may be implemented in different orders: either \texttt{TCHANGE} then branch, or branch then \texttt{TCHANGE}.
The actual sequence significantly affects the security and programming models, which effects we consider later.

To access the shared buffer \texttt{shm} and to permit execution of isolated code, the \gls{dpot} contains several records.
First, both \texttt{C\#1} and \texttt{C\#2} have their own records with read and write attributes.
POTIndex=0 with POIndex=1 enables RW access to data owned by \texttt{C\#1} for \texttt{C\#1} code.
POTIndex=1 with POIndex=2 enables analogous access for \texttt{C\#2}.
Access to the shared buffer is enabled via two separate \gls{dpot} records, each associated with the default POTIndex of the respective isolated components and with a dedicated POIndex=3 for the buffer: \texttt{C\#1} has write access with POTIndex=0 and POIndex=3, while \texttt{C\#2} has read access with POTIndex=1 and POIndex=3.

\gls{poe}2 is a complex extension comprising multiple tables, registers, and instructions; not all elements are considered in this example.
The scenario presented captures a canonical design pattern for isolated components: transitions between mutually distrusted parties and data exchange via a shared buffer.

\subsection{Challenges}

As the example shows, configuring \gls{poe}2 components is a complex task that is prone to error.
An incorrectly chosen index can grant unintended permissions or enable unintended transitions.
Below we summarise the key challenges of programming \gls{poe}2 and the motivation for a unified programming model.

\paragraph{Lack of Semantics}

\gls{poe}2 operates in a multi‑dimensional space of indices, where a particular permission is defined by a POTIndex defined by the base indices (POIndex and TIndex), the operation performed (read/write/\texttt{TCHANGE}/fetch), and the base indices of the target memory or state.
\gls{poe}1, Intel MPK, and other intra‑process isolation approaches, use a spatial index (i.e. a key) to denote a compartment -- protection domain for execution units.

In \gls{poe}2, the spatial index alone does not define an execution unit because the permission model is constructed from combinations of different indices.
\gls{poe}2 distinguishes permissions for code executing in memory with the same POIndex but different TIndexes, and it controls transitions between these states via the \gls{ttt}~\cite{arm_aarch64_s1poe2_2025}.

At the same time, permissions also change when the spatial index changes, and spatial changes are not explicitly controlled but are implicit: within a single TIndex nothing prevents branching between code located in different POIndexes (if pages are executable).
To control branching, a programmer must therefore assign different TIndexes.
In other words, \gls{poe}2 lacks a clear semantic for controlling transitions between states: it provides explicit control over only one of the two orthogonal characteristics.

\paragraph{Transition Risks}

The kernel’s asymmetric permission model for user–kernel interaction relies on system calls and trap‑based domain transitions.
When a user‑space thread issues a system call, it cannot influence how the kernel dispatches that call: execution traps to a well‑defined handler and user code has no mechanism to transfer control to arbitrary kernel addresses.

In contrast, in \gls{poe}2 operating at EL0\footnote{Other exception levels provide trap‑like domain transition mechanisms based on \texttt{TENTER} and \texttt{TEXIT}. This work considers \gls{poe}2 only in EL0.}, a caller may branch to any executable region of a callee’s memory if the branch is permitted (for example when both caller and callee share the same TIndex).
Conceptually, this implies that two mutually distrusted parties that communicate in an RPC‑like fashion cannot fully control the control‑flow of the interaction: the counterparty may transfer control to an arbitrary address within the permitted region.

The risk applies equally to interactions between trusted code and an untrusted sandbox.
A transition into the sandbox can be arranged by the trusted component, but control of the return or subsequent branches originating inside the sandbox is not constrained by the trusted component.
A sandboxed \gls{jit} can branch into code sequences within the trusted region that the trusted component did not intend to expose.
Control‑flow hardening primitives such as \texttt{BTI} or authenticated pointers do not eliminate this risk because they do not compel untrusted code to use specific trampolines or to restrict its branch targets to the addresses supplied by the trusted component.

These transition risks motivate a programming model that abstracts the multi‑index configuration and enforces controlled transition and permission semantics.

\section{Programming Model}\label{sec:program_model}

From the software perspective, \gls{poe}2 assumes a two‑phase life‑cycle for isolated software.
A privileged entity is responsible for deploying software components into distinct memory regions.
It then configures the \gls{poe}2 tables by assigning permissions to states defined by POIndexes and TIndexes.
In the second phase, the partitioned software executes and performs domain transitions without kernel involvement.

We develop our programming model to follow this system design.
The monitor is a privileged component within a process responsible for the life‑cycle of so‑called \emph{Complets}: a \gls{poe}2‑tailored abstraction for isolated binaries.
The Monitor deploys and links binaries to enable their communication.
It exposes a hostcall interface to \emph{Complets} and interacts with the kernel to configure \gls{poe}2 tables or to obtain mechanisms unavailable in user space. Conceptually, the model does not prevent \emph{Complets} from interacting with the kernel, but the performance benefits of \gls{poe}2 derive from enabling intra‑process communication without kernel involvement.

\paragraph{Security assumptions}

By default, we assume compartmentalised code is adversarial.  
It may execute any instruction; we assume it is hostile and aims to escape its compartment.  
Management software is trusted: its security‑critical components (in particular, trampolines) are assumed implemented correctly and free of bugs, while the remainder of a binary may contain ROP gadgets or other payloads that compartmentalised code may attempt to exploit.  
The kernel is trusted and \gls{poe}2 is assumed to behave according to its specification.  
Less strict security models can be derived and implemented using the programming model presented here.

\begin{table*}[h]
\centering
\resizebox{\textwidth}{!}{
\begin{tabular}{l p{8cm}}
	\hline
	\textbf{API Functions}                                       & \textbf{Purpose / Details}                               \\ \hline
	\texttt{mon = init\_default\_complet()}                       & Initialise POE2 tables for Monitor code                  \\
	\texttt{c1 = deploy\_complet(path, base\_addr, size)}         & Allocate memory, load elf inside, setup POE2 tables      \\
	\texttt{t1 = deploy\_complet("trampoline", base\_addr, size)} &                                                          \\
	\texttt{enable\_transition(c1, t1)}                          & Update POE2 tables to enable transitions from c1 into t1 \\
	\texttt{enable\_transition(t1, mon)}                         &                                                          \\
	\texttt{install\_hook(c1, "write", t1)}                      & Replace "write" function with trampoline provided by t1  \\
	\texttt{c1\_ctx,mon\_ctx = alloc\_shared(size)}              & Allocate memory to store thread contexts                 \\
	\texttt{enable\_shared(c1\_ctx, t1, rw) }                    & Grant read-write access to shared memory for context     \\ \hline
	                                                             &
\end{tabular}
}
\caption{API to use complets by Monitor: create complets for code and trampoline, bridge complet with monitor via trampoline}
\label{tab:api}
\end{table*}

\paragraph{Complet}
We introduce a system abstraction, the \emph{complet} — a \emph{compartmentalised applet}.
\emph{Complets} are mutually distrusted, cannot access each other’s memory, and interact only via shared memory.
Each complet defines a protection domain for its threads; threads perform transitions between complets in a trusted manner to realise inter‑complet communication.

Under the default, strict security model, complets are mapped to \gls{poe}2 structures as follows: 
\begin{itemize}
	\item no two complets share the same TIndex or POIndex. Consequently, allocating a new complet requires assigning fresh POIndex and TIndex values
	\item Shared memory regions receive distinct POIndexes. Each complet obtains its own access rights to shared regions via a POTIndex derived from its POIndex and TIndex.
\end{itemize}

\subsection{Trampolines}

Threads within \emph{complets} cannot branch into addresses belonging to other \emph{complets}, yet they must exchange information.
Domain transitions between \emph{complets} are provided by the compartmentalisation runtime as trusted trampolines.
The programming model is centred on the privileged Monitor, which mediates and bridges isolated components according to a predefined partitioning.

The partitioning may be encoded in the ELF image or established by the Monitor at deployment; in either case the model assumes that isolated code is \gls{poe}2‑protected and isolation‑unaware, and therefore does not advertise public interfaces or interact with the Monitor autonomously.
The Monitor performs \gls{poe}2 configuration, exposes a hostcall interface to \emph{complets} installing hooks, and interacts with the kernel on their behalf.
This design preserves the performance advantages of \gls{poe}2 by enabling intra‑process communication without routine kernel mediation.

\section{Implementation}\label{sec:impl}
Table~\ref{tab:api} shows API functions implementing basic usage of complets.
The code initialises three complets: compartmentalised code with binary pointed by \texttt{*path}, complet with trusted code of trampoline, and initialises the default complet for Monitor code.
It than enables transitions between complets updating the \gls{dpot} and \gls{ttt} for the corresponding indexes stored inside complet objects.
Subsequently, \texttt{install\_hook()} patches the code inside complet to route the \texttt{write()} function into trusted trampoline.
Finally, the code allocates memory to store component contexts and grants permissions to these contexts.

\begin{figure}[t!]
	\centering
	\includegraphics[width=0.4\textwidth]{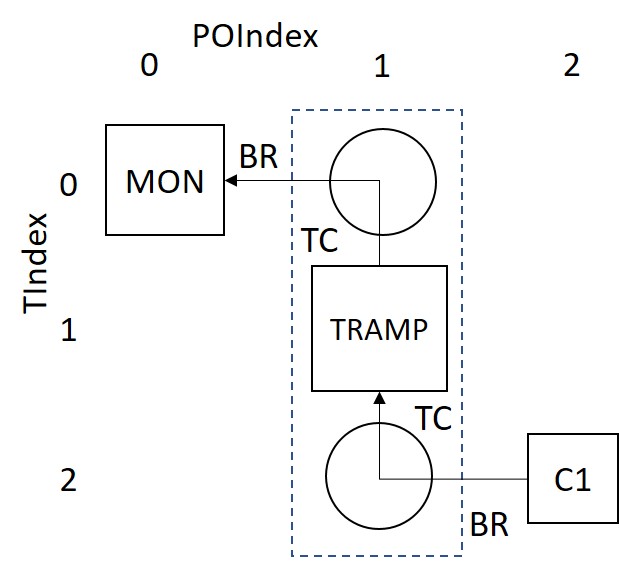}
	\caption{Transition from complet to Monitor via trusted trampolines: preserves ID and guarantees secure landing }
	\label{fig:tramps}
\end{figure}

Figure~\ref{fig:tramps} shows the secure transition enforced by the programming model.
The transition begins inside a complet: the hook installed by the Monitor inserts a branch instruction that transfers control into the memory region of TRAMP.
In this state, the compartmentalised code may perform only two operations: execute its own code (i.e. execute in 
[POIndex=2, TIndex=2]), or branch into a trampoline's POIndex.
The trampoline code is minimal and stateless. This design prevents payload extraction and relies on a small code size to reduce the attack surface; the trampoline is below 100 assembler instructions.
While in the [POIndex=1,TIndex=2] state, execution is restricted to either issuing a \texttt{TChange} instruction with transition into the [POIndex=1,TIndex=1] state, or performing a jump that returns control to C1.
The code may be further constrained using BTI instructions to enforce landing at the trampoline entry.
After issuing \texttt{TChange}, the trampoline obtains the TIndex of the originating complet, which can subsequently be used to identify the caller.

The core trampoline implements a context switch from a complet's context to the Monitor context using a memory buffers allocated by the Monitor.
These buffers are writable only when the trampoline is in the [POIndex=1,TIndex=1] state.
The trampoline then escalates its TIndex to the Monitor’s value and branches into the Monitor code responsible for handling the hook for the \texttt{write} function.

\subsection{Evaluation of Key Design Patterns}\label{sec:design_patterns}

We implement a prototype of the compartmentalisation system, the Monitor, and the trusted trampoline.
The full one-way transition requires approximately 75 instructions to store and load contexts and to perform domain transitions.
We add support for \gls{poe}2 in the Linux kernel by extending the existing \gls{mpk} API with new syscalls; this change comprises roughly 300 new source lines of code for \gls{poe}2 table management.
Using the compartmentalisation library, we prototype representative design patterns for intra-process isolation and evaluate them on a POE2-enabled functional model (Arm's FVP~\cite{arm_fvp}).

\paragraph{Symmetric}

The symmetric isolation scheme assumes mutually distrusted parties that cannot directly access each other and require a third party for data exchange or transitions.
Prior systems that follow this pattern include Hodor~\cite{hodor}, IskiOS~\cite{iskios}, libmpk~\cite{libmpk}, SOAAP~\cite{soaap}, LwCs~\cite{lwc}, SeCage~\cite{secage}, CubicleOS~\cite{cubicleos}, FlexOS~\cite{flexos}, and others.
The complet's API and programming model adopt this pattern by default; therefore, systems that use symmetric partitioning can be reproduced with complets and \gls{poe}2 with minimal adaptation.

\paragraph{Asymmetric}

The asymmetric isolation scheme for \gls{jit}ed code separates code into two parts: an untrusted region for generated code and a trusted one.
The untrusted region is constrained to its own memory, while the trusted region has access to both its own memory and the untrusted region’s memory.
This arrangement enables efficient data exchange because the trusted code can access data directly without copying between protection domains.
Systems that employ this pattern include ERIM~\cite{erim}, Shreds~\cite{shreds}, CETIS~\cite{cetis}, E-IM~\cite{eim}, PrivTrans~\cite{privtrans}, Wedge~\cite{wedge}, and others.

The Complets API supports this scenario: the trusted component (Monitor) can request read/write access to an untrusted complet in the same manner as requesting access to shared memory.
Because the Monitor’s and the untrusted complet’s key indexes are known, the asymmetric model is straightforward to implement by enabling read/write entries in the \gls{dpot} for the Monitor’s POTIndex.

\paragraph{Enclave}

Secure enclaves, as introduced by Intel SGX~\cite{sgx14}, represent an asymmetric design in which an intra-process isolated component has full access to the host process (except other enclaves), while non-enclaved software has no access to enclaves.
Enclave deployment and manipulation use special instructions and are not controlled by the \gls{mmu} in the same way as \gls{mpk} or \gls{poe}.
Despite differences in lifecycle and security assumptions relative to \gls{poe}2, the enclave model for data access can be reproduced on top of \gls{poe}2 and the Complets API.
Specifically, the Monitor can restrict its read–write access to designated “enclave” complet and enable read–write access to its own memory for those complets.

\section{Discussion}
\paragraph{Weaker Security}

The Complets API supports a strong security model in which isolated software is treated as hostile.
To ensure transitions from isolated code land at intended Monitor functions, we use an intermediate trusted trampoline that is stateless and has a small footprint.
This approach enforces trusted transitions but requires additional POIndex/TIndex values and the use of \texttt{TChange} instructions, which may incur performance overhead.
A weaker programming model is possible: if the Monitor code is assumed bug-free and incapable of containing payload, isolated code may transfer control directly into Monitor code.
The total number of registers stored and restored during the transition, which is one of source of performance overhead of isolation, can also be decreased.
Even in this weaker model, we recommend using BTI instructions to enforce landing at intended entry points; this design can be realised using the Complets API.

\paragraph{Why Complets}

We introduce the abstraction of \emph{complets} rather than reuse the term \emph{compartments}, whilst use \emph{compartmentalisation} as a technique.
We view compartments as primarily associated with spatial characteristics of code, i.e., the protection key tied to particular memory (POIndex in \gls{poe} terminology).
The complet abstraction combines spatial and temporal characteristics and hides those details behind a single interface.
Thus, while the term compartment remains applicable as a general concept, complets represent a distinct, richer abstraction that can be partially mapped to compartments.

\paragraph{Future Work}

The performance characteristics of POE2 depend strongly on microarchitectural implementation: new tables whose values must be cached, new instructions that operate on cached values, new registers, and related mechanisms.
Consequently, while developing a compartmentalisation library that implements complets using \gls{poe}1 is a reasonable proof of concept, its performance hardly can be extrapolated to \gls{poe}2.
Future work will implement \gls{poe}2 on a suitable platform for performance evaluation and will exercise the proposed model with real-world scenarios.

\bibliographystyle{plain}
\bibliography{sample-base}

\end{document}